\documentclass[]{spie}  

\usepackage{amsmath,amsfonts,amssymb}
\usepackage{graphicx}
\usepackage[colorlinks=true, allcolors=blue]{hyperref}

\title{The Fast Atmospheric Self-Coherent Camera Technique: Laboratory Results and Future Directions}

\author[a,b]{Benjamin L. Gerard}
\author[b,a]{Christian Marois}
\author[c]{Rapha\"el Galicher}
\author[c]{Pierre Baudoz}
\author[d]{Polychronis Patapis}
\author[e, d]{Jonas K\"uhn}

\affil[a]{University of Victoria, Department of Physics and Astronomy, 3800 Finnerty Rd, 
Victoria, V8P 5C2, Canada}
\affil[b]{National Research Council of Canada, Astronomy \& Astrophysics Program \\
5071 West Saanich Rd\\
Victoria, V9E 2E7, Canada}
\affil[c]{Lesia, Observatoire de Paris, PSL Research University, CNRS, Sorbonne Universit\'es, Univ. Paris Diderot \\
UPMC Univ. Paris 06, Sorbonne Paris Cit\'e, 5 place Jules Janssen \\
92190 Meudon, France}
\affil[d]{ETH Zurich, Institute of Particle Physics and Astrophysics, Wolfgang-Pauli-Strasse 27, CH-8093 Zurich, Switzerland}
\affil[e]{University of Bern, Center for Space and Habitability, Gesellschaftsstrasse 6, 3012 Bern, Switzerland}

\authorinfo{Further author information: Send correspondence to Benjamin L. Gerard (bgerard@uvic.ca)}

\begin{document} 
\maketitle

\begin{abstract}
Direct detection and detailed characterization of exoplanets using extreme adaptive optics (ExAO) is a key science goal of future extremely large telescopes (ELTs). However, wavefront errors will limit the sensitivity of this endeavor. Limitations for ground-based telescopes arise from both quasi-static and residual AO-corrected atmospheric wavefront errors, the latter of which generates short-lived aberrations that will average into a halo over a long exposure. We have developed and tested the framework for a solution to both of these problems using the self-coherent camera (SCC), to be applied to ground-based telescopes, called the Fast Atmospheric SCC Technique (FAST). In this paper we present updates of new and ongoing work for FAST, both in numerical simulation and in the laboratory. We first present numerical simulations that illustrate the scientific potential of FAST, including, with current 10-m telescopes, the direct detection of exoplanets reflected light and exo-Jupiters in thermal emission and, with future ELTs, the detection of habitable exoplanets. In the laboratory, we present the first characterizations of our proposed, and now fabricated, coronagraphic masks.
\end{abstract}


\section{INTRODUCTION}
\label{sec:intro}  
Detection and characterization of habitable exoplanets is a key science goal of the astronomical and broader community. Although extremely large telescopes (ELTs) will reach the resolution needed to directly image habitable exoplanets around nearby M stars in reflected light \cite{guyon_hab} and FGK stars in thermal emission \cite{quanz_hab}, the deepest contrast limits reached with current instruments on 8 m-class telescopes\cite{me_gpi} are still orders of magnitude higher that what would be needed to detect such habitable exoplanets on ELTs. 

The main error terms preventing current instruments from reaching deeper contrast limits close on the star, where habitable exoplanets will be on ELTs, are generated from differential temporal and chromatic errors, including both residual adaptive optics (AO) and quasi-static aberration\cite{ao,ncpa}, as well as from residual diffraction left unattenuated by an imperfect coronagraph\cite{guyon_coron}. In other words, if an ELT instrument were perfectly stable and/or achromatic, averaging AO residuals over time could in principle enable reaching the desired contrasts needed to directly image habitable exoplanets. Instead of relying on this approach of stability and achromaticity, in Gerard et al. (2018a\cite{fast} and 2018b\cite{fast_spie}; here after G1 and G2, respectively) we proposed an alternative method using the Self-Coherent Camera (SCC)\cite{scc_orig}, called the Fast Atmospheric SCC Technique (FAST). FAST relies on the method of recording coronagraphic images on millisecond timescales, freezing the dynamic nature of temporal aberrations in individual frames and in doing so removing this effect as a fundamental limitation in reaching deeper contrasts. Note that another similar approach using millisecond telemetry from both coronagraphic images and a synchronized adaptive optics system has been proposed\cite{frazin,frazin_algo}. Although we have not yet presented a dedicated FAST solution to chromaticity, the SCC method relies on the monochromatic principle of coherence, distinguishing self-coherent stellar aberrations in the coronagraphic image from an incoherent exoplanet. FAST is therefore operational without requiring a large bandpass and not limited by chromaticity, although a more detailed broadband solution will be presented in a later paper. 

\begin{figure}[!ht]
\centering
\includegraphics[width=1.0\textwidth]{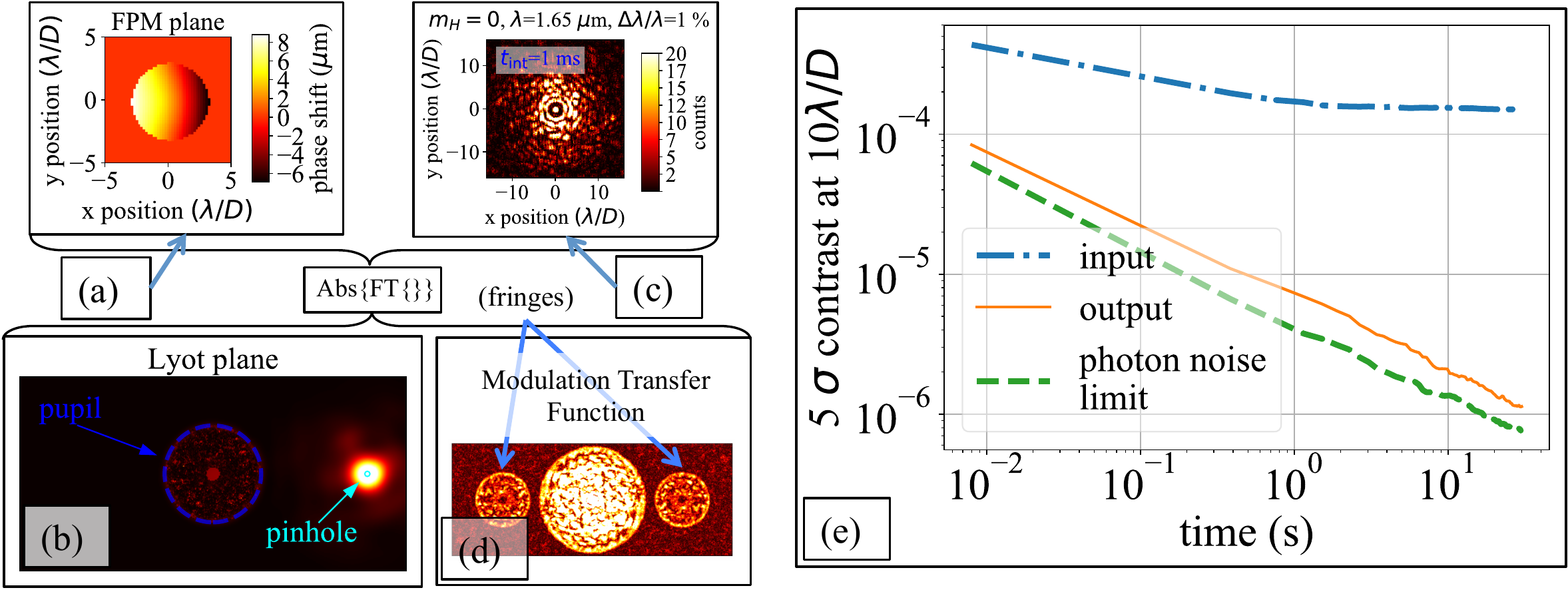}
\caption{Summary of FAST, adapted from Gerard et al. 2018a\cite{fast}, 2018b\cite{fast_spie}. A new focal plane mask (FPM; a) is designed to concentrate light into the off-axis pinhole of the Lyot plane (b; a Lyot stop then blocks the light outside of the pupil and pinhole from being transmitted to the coronagraphic image). As a result of this design, fringes in the coronagraphic image (c and d), generated from optical interference between the Lyot plane pinhole and pupil, are detected in a millisecond exposure, even when only detecting a few photons. (e) Stacking subtracted millisecond frames continually improves contrast with integration time, reaching close to the photon noise limit. FAST Speckle subtraction algorithms rely on fringe detection in millisecond exposures and can be implemented in post-processing\cite{fast} and/or live AO control\cite{fast_spie}.}
\label{fig: fast_sum}
\end{figure}

Figure \ref{fig: fast_sum} summarizes FAST. 
In short, a new coronagraphic mask allows for SCC fringe detection on millisecond-timescales. This fringe detection enables subtraction algorithms to be photon noise-limited, no longer limited by quasi-static speckles and AO residuals. In principle, the gap limiting current instruments to detecting habitable exoplanets on ELTs can be achieved with FAST simply by integrating for long enough. However, further development of FAST is needed to validate this principle. In this paper we present further development of ongoing FAST work, including numerical simulations of science cases using FAST (\S\ref{sec: science} and the first laboratory tests of fabricated coronagraphic masks (\S\ref{sec: lab}).
\section{Scientific Motivation}
\label{sec: science}
\begin{figure}[!h]
\centering
\includegraphics[width=1.0\textwidth]{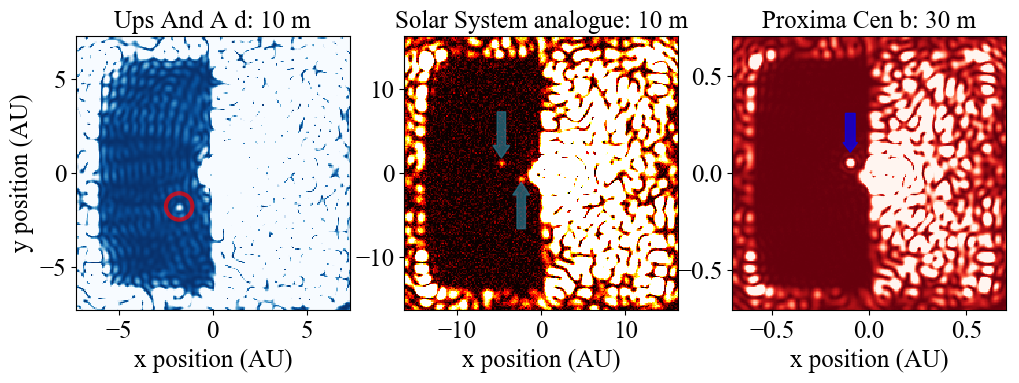}
\caption{Simulated optimistic FAST science cases for current 10 m- (left and middle) and future 30 m- (right) class telescopes. End-to-end simulations will be presented in a future publication. Left: A 0.85 $\mu$m, 5 hour exposure of the radial velocity exoplanet Ups And A d (10 $M_\text{Jup}$, 2.5 AU at 13.5pc; Butler et al. 1997\cite{ups_and}). Middle: A 1.65 $\mu$m, 1 hour exposure of two 1 $M_\text{Jup}$ planets at 5 and 10 AU orbiting a 30 Myr star at 30 pc. Right: A 1.22 $\mu$m, 30 minute exposure of the radial velocity exoplanet Proxima Cen b (1-3 Earth masses in the habitable zone at 1.3pc; Anglada-Escud{\'e} et al. 2016\cite{proxima_cen}).}
\label{fig: fast_science}
\end{figure}

The improvements from fast focal plane wavefront sensing methods, such as FAST, could enable new science on current 10-m class telescopes and future ELTs, as illustrated in Fig. \ref{fig: fast_science}. Note that these simulations assume DM correction by a factor of 50, in contrast, over the full half dark hole and continual subtraction by post-processing down to the photon noise limit as a function of exposure time (see below for more details). While probably optimistic, the simulations show that fast focal plane wavefront sensing could enable the first direct images of indirectly detected exoplanets. Although host star radial velocities have revealed a plethora of such giant exoplanets and measured their masses, directly imaging their reflected starlight would provide an atmospheric characterization and crucial dynamical mass calibration (since mass can only be inferred through imaging). Such detections would also serve as a technology demonstration of the tolerances and requirements for reflected light imaging of habitable exoplanets on ELTs. Fast near-infrared focal plane wavefront control with current facilities could also detect thermal emission from young ($\lesssim100$ Myr), self-luminous exoplanets down to 1 $M_\text{Jup}$ at 5 AU around 40 targets (E. Neilsen, private comm.). Since current direct imaging surveys are not yet sensitive to these Solar System scales (Neilsen et al. 2019)\cite{gpies}, fast focal plane wavefront sensing methods, such as FAST, could provide the first estimates of the commonality of Jupiter ``twins'' beyond our Solar System. As a result of the tests and calibrations with giant exoplanets, once the ELTs come online in the late 2020s, methods such as FAST could enable the direct detection of exoplanets in the habitable zone around the nearest stars.

The simulations used to generate Fig. \ref{fig: fast_science} are run as follows. All simulations use our newly-proposed Tip/tilt Gaussian Vortex (TGV) coronagraph\cite{tgv}, updating the previous focal plane mask (FPM) design from G1 to enable better diffraction suppression and inner working angle ($\sim$2 lambda/D) while still maintaining sufficient fringe visibility for FAST to operate on millisecond timescales. As in G1, speckles are generated from an assumed 100 nm rms residual AO error, 25 nm rms static error, and 1\% rms static intensity error, all normalized in the entrance pupil. The bandpass central wavelength for each panel is assumed to scale the aforementioned wavefront error, and we assume use of the full bandpass to count the number of photons reaching the coronagraphic image (from left to right in Fig. \ref{fig: fast_science}: I band, H band, and J band). However, a monochromatic Fraunhofer propagation at the center of each bandpass is otherwise used to simulate an image. Although with this approach we do not include effects of wavefront chromaticity, focal plane mask chromaticity, and/or fringe smearing with wavelength, the point of these simulations is not to demonstrate end-to-end broadband performance. Instead, we use these simulations to illustrate the potential science cases that fast focal plane wavefront sensing methods, such as fast FAST, could optimistically enable. With this in mind, we assume that the monochromatic performance gain simulated in G1 (i.e., continually reaching the photon noise limit with increasing exposure time) will be the same in our broadband setup here; a separate forthcoming paper will address these assumptions with detailed end-to-end broadband simulations. Transmission through the atmosphere, instrument throughput, and detector quantum efficiency are assumed to be 90\%, 20 \%, and 60\%, respectively. A 10 or 30 m diameter unobscured circular telescope pupil is assumed, titled in each panel of Fig. \ref{fig: fast_science}. Sky background levels for each filter are obtained from Keck/NIRC2 measurements\footnote{\url{https://www2.keck.hawaii.edu/inst/nirc2/filters.html}}. The middle panel in Fig. \ref{fig: fast_science} assumes a $m_H=5$ host star; the cooling curves from Spiegel \& Burrows (2012) \cite{cool_curve} are used to calculate the exoplanet H band magnitude, assuming a hot start (highest entropy) formation scenario and a 30 Myr age. Ups And A b has an assumed planet-to-host star astrophysical flux ratio of $10^{-8}$, assuming a cloudy atmosphere without water, phase function of 0.5 at 60 degrees, and geometric albedo of 0.7 (M. Marley, private communication). Proxima Cen b has an assumed planet-to-host star astrophysical flux ratio of $8\times10^{-7}$ (Guyon et al. 2011)\cite{guyon_hab}. As in G1 we assume that the photon noise limit can be reached by post-processing, and as in G2 we assume that this limit can be decreased further by deformable mirror (DM) control. Thus, the assumed exposure times in Fig. \ref{fig: fast_science} are used to determine the photon-noise limit image, calculated from the difference between a propagation with and without photon noise. The propagation only uses a single complex wavefront realization and does not include a full translating residual atmospheric phase screen as in G1 or G2; again, here we are simply illustrating the potential advantage of FAST in a simple model. The photon noise limit image is then divided by $\sqrt{50}$, assuming that DM control improves contrast by a factor of 50 at all image separations within the dark hole. The coronagraphic image for the on-axis propagated wavefront (with photon noise) is then normalized to the contrast of the aforementioned photon noise limit image. Contrast is computed as the standard deviation over the half dark hole of the DM control region. The propagated off-axis wavefront (also with photon noise), scaled to the aforementioned planet-to-star ratio, is then added to the scaled on-axis star in the coronagraphic image plane and displayed in Fig. \ref{fig: fast_science}.

\section{Laboratory Tests}
\label{sec: lab}
In this section we present the first laboratory test of FAST, including results from the LESIA THD2\cite{thd2} (\S\ref{sec: thd} and ETH Zurich\cite{eth} (\S\ref{sec: eth}) high contrast imaging testbeds.
\subsection{First laboratory Tests at LESIA}
\label{sec: thd}
\begin{figure}[!h]
\centering
	\begin{minipage}[b]{0.2\textwidth}
		\begin{center}
		\includegraphics[width=1.0\textwidth]{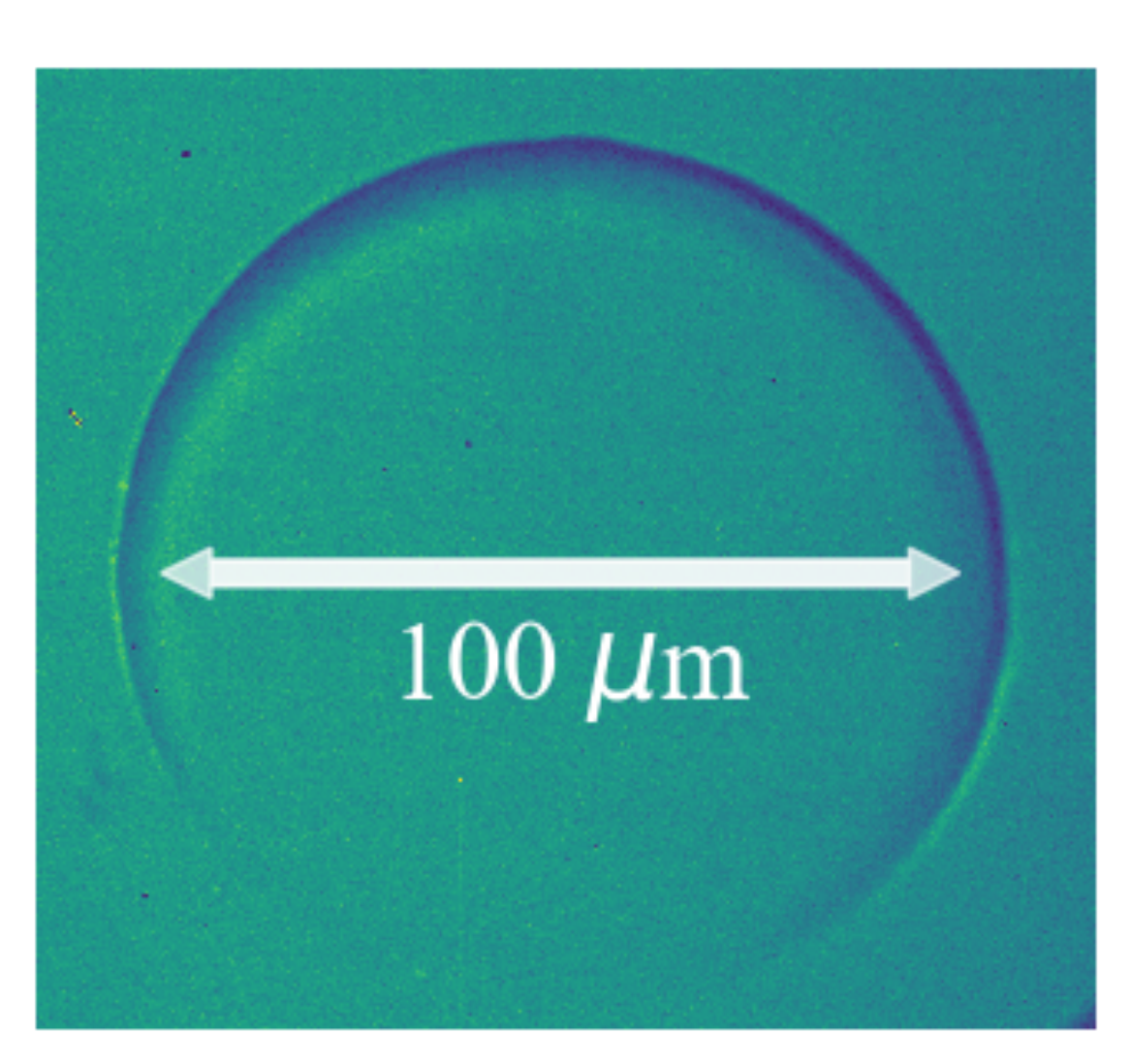}
		(a)
		\end{center}
	\end{minipage}
	\begin{minipage}[b]{0.4\textwidth}
		\begin{center}
		\includegraphics[width=1.0\textwidth]{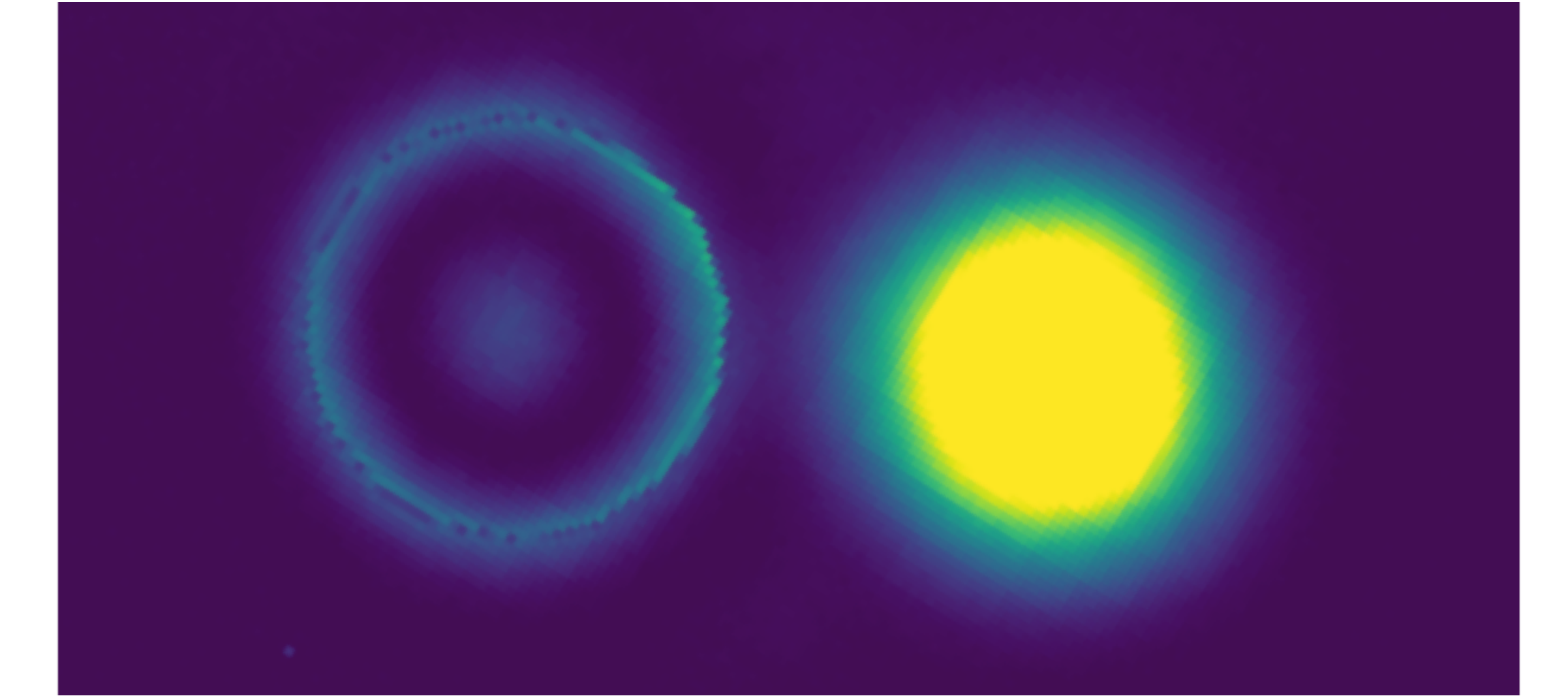}
		(b)
		\end{center}
	\end{minipage}
	\begin{minipage}[b]{0.205\textwidth}
		\begin{center}
		\includegraphics[width=1.0\textwidth]{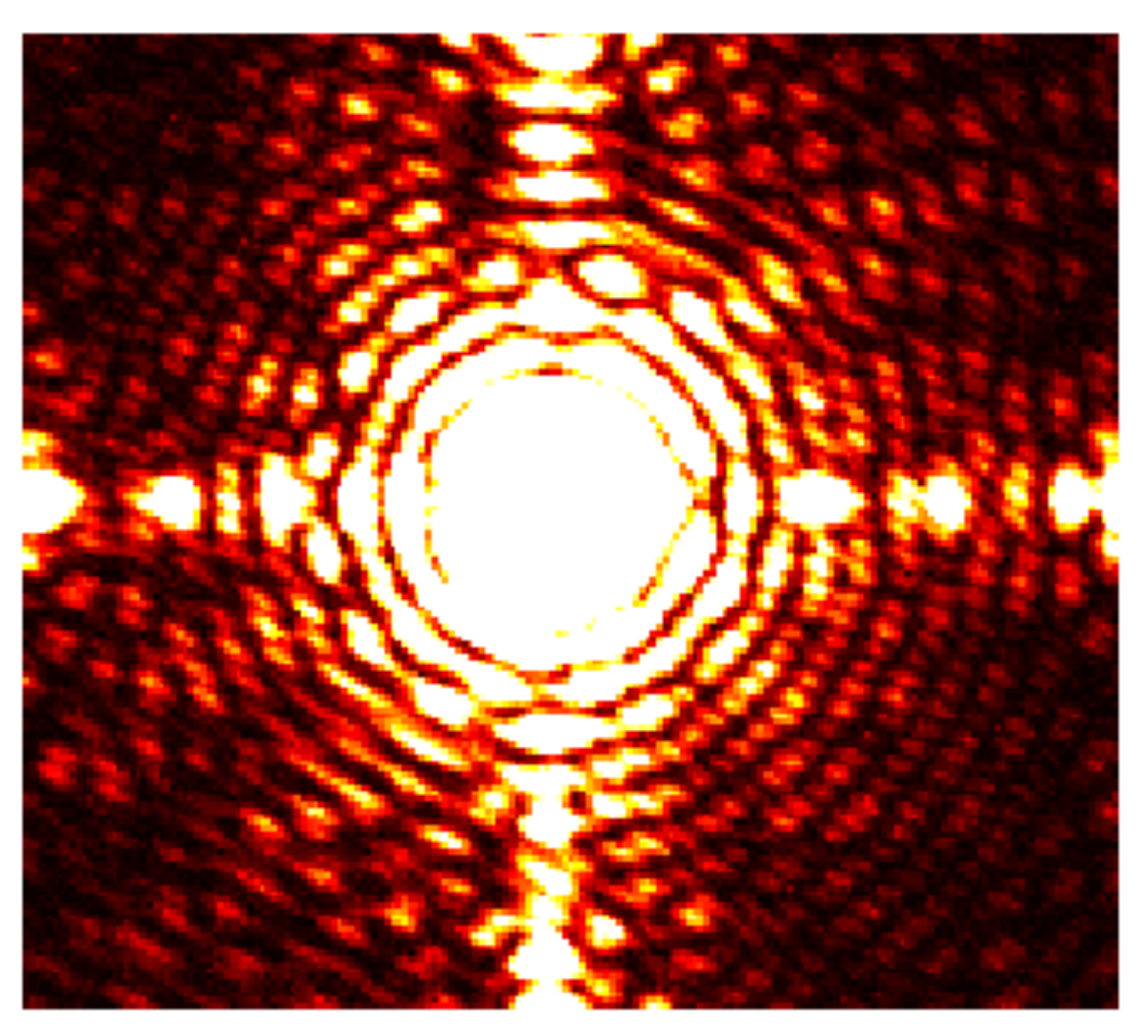}
		(c)
		\end{center}
	\end{minipage}	
	\begin{minipage}[b]{0.45\textwidth}
		\begin{center}
		\includegraphics[width=1.0\textwidth]{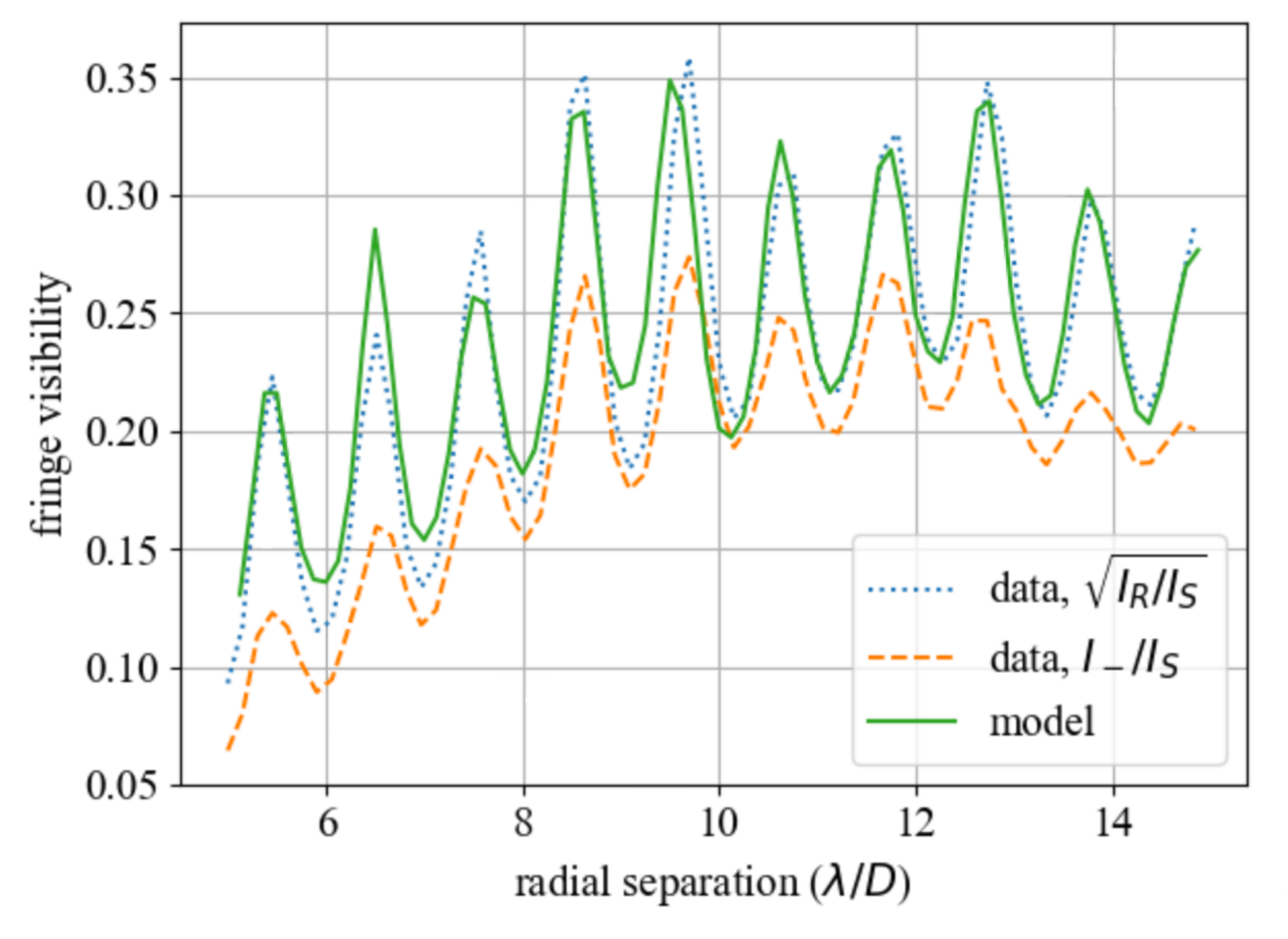}
		(d)
		\end{center}
	\end{minipage}	
	\begin{minipage}[b]{0.45\textwidth}
		\begin{center}
		\includegraphics[width=1.0\textwidth]{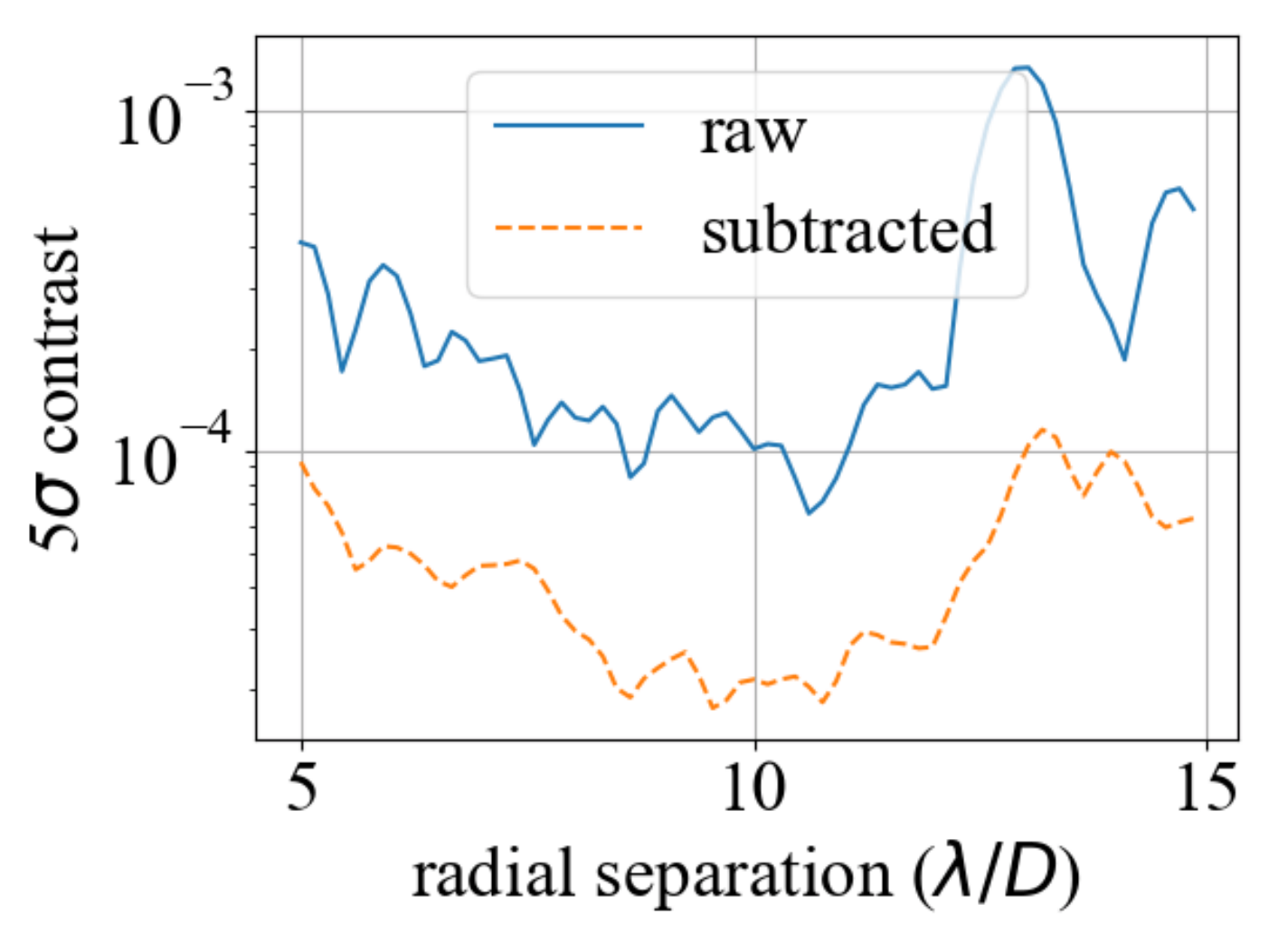}
		(e)
		\end{center}
	\end{minipage}	
\caption{Laboratory tests at LESIA. (a) Microscope image of the fabricated TG FPM, (b) Lyot pupil image, (c) SCC image, (d) fringe visibility curves (measuring the amplitude of the fringes relative to the speckles in the coronagraphic image), and (e) contrast curves showing the image from (c) (i.e., raw) and our best subtraction.}
\label{fig: fvs}
\end{figure}

A transmissive coronagraphic mask was designed for operation at 700 nm on the THD2 bench and fabricated by Zeiss Optics. Fabrication was done by ion etching into a SiO$_2$ wafer, followed by anti-reflective coating for 700 nm. Eleven different samples were all fabricated from the same wafer; we used microscope images to identify the best mask by inspection, shown in Figure \ref{fig: fvs} (a). Due to constraints set by other mask designs also being fabricated on the same wafer, the peak-to-valley optical path difference was limited to 2$\pi$ radians; using the tip/tilt+Gaussian (TG) design from G1 and the position of the off-axis Lyot pinhole on the THD2 bench, the inner working angle (IWA) of the mask was limited to 0.7 $\lambda/D$. With no apodizer and/or phase mask design outside this central 1.4 $\lambda/D$ diameter TG region, our goals here, rather than to demonstrate a deeply corrected contrast, were to provide an initial measurement and characterization of fringe visibility in the pupil and focal planes downstream of the aligned FPM. With this in mind, Fig. \ref{fig: fvs} (b) - (e) shows a compilation of FAST results from the LESIA high contrast testbeds, obtained in March 2019.

Initial testing at LESIA was carried out in the ``Corono Lab''\cite{thd}, formerly the original THD lab and now used for initial inspection, alignment, and testing of coronagraphic masks before transferring them to the main THD2 lab. After microscope inspection and mask alignment on the bench, the Lyot plane pupil image using this mask is shown in Fig. \ref{fig: fvs} (b); the THD2 bench cannot be configured to take analogous coronagraphic pupil images, with the FPM aligned and Lyot stop removed (i.e., taken on the same camera as the focal plane images using a deployable lens to re-image the pupil), although coronagraphic pupil images can still be acquired with the Lyot stop in\cite{thd2}. The Corono lab uses a standard Lyot stop coronagraph design with no off-axis pinhole, and so coronagraphic images with SCC fringes can only be obtained on the THD2 bench. Note that the f number in the FPM plane is 30\% smaller in the Corono lab than in the THD2 lab (i.e., providing a TG FPM tilt angle that is 30\% shallower and an IWA than is 30\% larger). Assuming an off-axis Lyot stop pinhole lies at the peak intensity of the off-axis pupil in Fig. \ref{fig: fvs} (b), we measure a rough integrated fringe visibility (i.e., cumulative of the intensity in the pinhole divided by cumulative intensity in the central pupil) of 4\%, consistent with simulations that account for the mask design and bench setup.

Next, bringing the same mask to the THD2 lab produced Figures \ref{fig: fvs} (c) - (e). In Fig. \ref{fig: fvs} (d) we calculate fringe visibility as a function of separation by recording the fringed SCC image (Fig. \ref{fig: fvs} c), the coronagraphic image with the off-axis Lyot pinhole closed ($I_S$), and the pinhole point spread function (PSF) image with the central Lyot pupil blocked ($I_R$). With the these three images, fringe visibility can be calculated in two different ways, both of which are shown in Fig. \ref{fig: fvs} (c). One way uses the two unfringed images ($I_S$ and $I_R$), while the other uses the fringed image and the unfringed coronagraphic image ($I_S$); the latter isolates the fringe amplitude in the Fourier plane of the SCC image (the optical transfer function, or OTF) to generate the standard $I_-$ quantity used in Baudoz et al. (2006)\cite{scc_orig} and subsequent papers. Mathematically and numerically, these two methods should yield exactly the same curve; observationally they do not. We will return to a discussion of this discrepancy in the following paragraph. Finally, 5$\sigma$ contrast curves are shown in Fig. \ref{fig: fvs} (e) for the raw input image (Fig. \ref{fig: fvs} c) and our best subtraction (using a direct measurement of the pinhole PSF on the same wavefront realization as in G1), illustrating that contrast gain is limited to a factor of about 5 at all separations.

The main factor causing the discrepancies in Fig. \ref{fig: fvs} (d) and subtracted contrast limitations in Fig. \ref{fig: fvs} (e) arise from our coronagraph design and limited dynamic range in the coronagraphic image. As described above, because our IWA was limited to 0.7 $\lambda/D$, the center of the image in Fig. \ref{fig: fvs} (c) is about 300 times brighter than speckles at mid to high separations. As a result, to prevent saturation in the PSF core (filtering saturated SCC images in the Fourier domain can cause significant systematic effects, as discussed below), speckles from 5 to 15 $\lambda/D$ are only detected at about a hundred counts above the read noise of the Andor s-CMOS detector\cite{thd2}. Thus, the marginal gain in Fig. \ref{fig: fvs} (e) is actually reaching the background noise limit; in order to go deeper, we would need a better coronagraph design and/or a higher dynamic range detector. In future laboratory demonstrations and papers we will use the TGV design, which rejects much more light outside of the central pupil in the Lyot plane and should therefore significantly improve over these current limitations. Similarly, in Fig. \ref{fig: fvs} (d), the discrepancy between the two measured fringe visibility curves are attributed to the same ``low signal-to-noise ratio (S/N) effect.'' Photon noise from the bright central core of the coronagraphic image is about 10 times higher than noise at mid to high spatial frequencies; as a result, the side lobes in the image OTF are detected at a S/N about 10 times lower than it could be with a better coronagraph. Instead, an algorithmic approach is to use a Fourier filter in the image plane to attenuate this effect (typically a Butterworth filter), masking the central star and it's associated photon noise. However, systematic effects, trading off between decreasing photon noise from the central star and decreasing signal from the fringes, still limit this approach. The orange dashed curve in Fig \ref{fig: fvs} (d) uses a particular set of Butterworth parameters to highlight this systematic discrepancy.

\subsection{ETH Zurich Tests}
\label{sec: eth}
\begin{figure}[!h]
\centering
	\begin{minipage}[b]{0.42\textwidth}
		\begin{center}
		\includegraphics[width=1.0\textwidth]{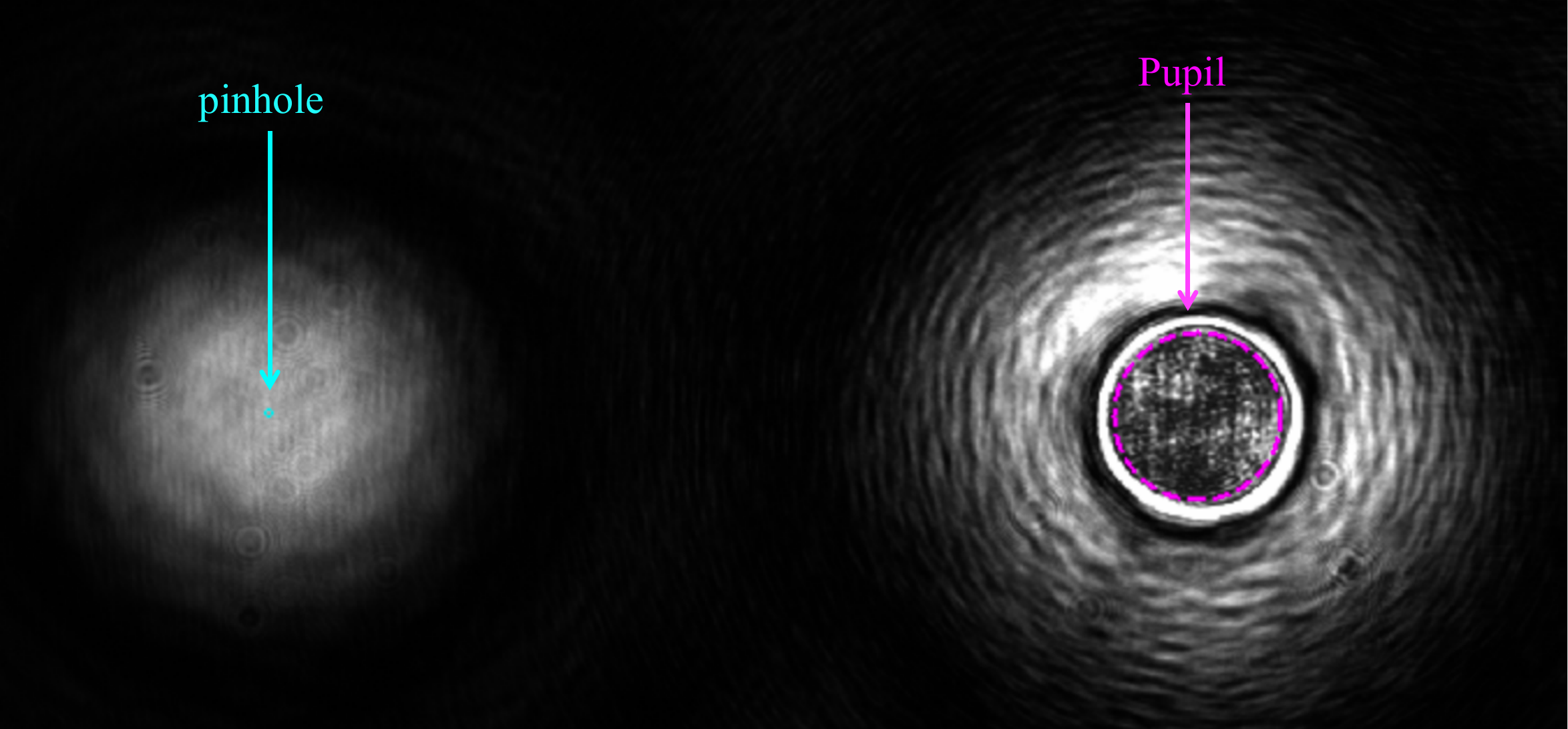}
		(a)
		\end{center}
	\end{minipage}
	\begin{minipage}[b]{0.215\textwidth}
		\begin{center}
		\includegraphics[width=1.0\textwidth]{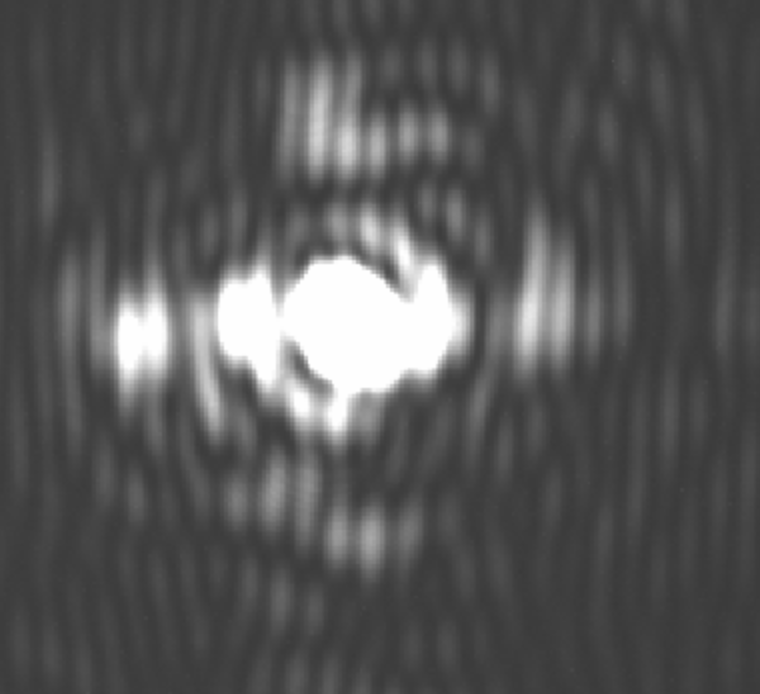}
		(b)
		\end{center}
	\end{minipage}
	\begin{minipage}[b]{0.34\textwidth}
		\begin{center}
		\includegraphics[width=1.0\textwidth]{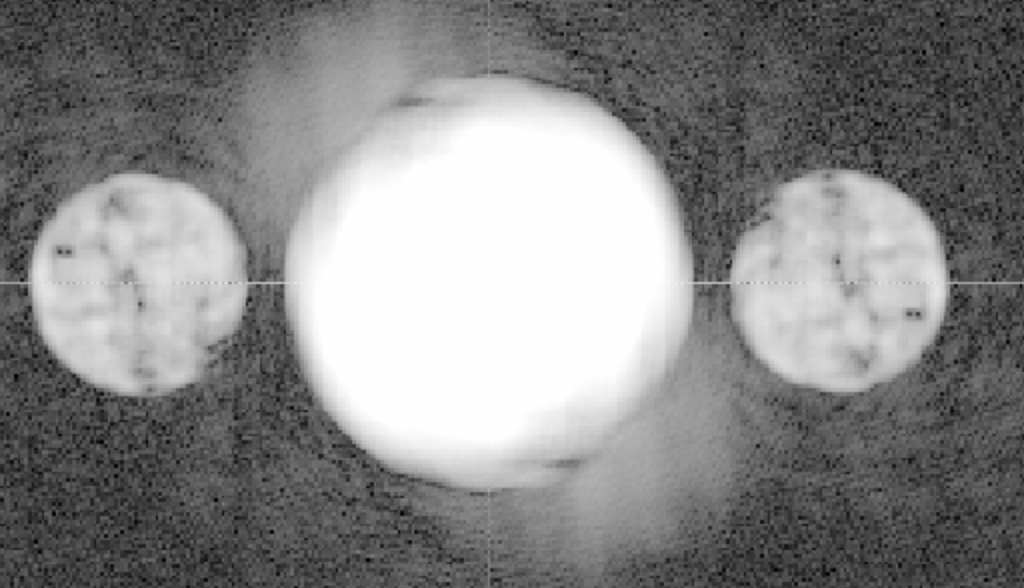}
		(c)
		\end{center}
	\end{minipage}	
\caption{Laboratory images from the ETH Zurich High Contrast Imaging Lab\cite{eth}. (a) Lyot plane intensity, before a Lyot stop is applied. (b) SCC coronagraphic Image. (c) modulation transfer function of (b).}
\label{fig: eth}
\end{figure}
We ran additional tests of our new TGV design using the ETZ Zurich high contrast imaging lab\cite{eth}, which includes an adaptable coronagraphic setup using a liquid crystal on-silicon spatial light modulator (SLM) in the focal plane. We applied the TGV design to the SLM to produce the images in Fig. \ref{fig: eth}. Using the illustrated ``pupil'' and ``pinhole'' apertures in Fig. \ref{fig: eth} (a), we measure and integrated Lyot plane fringe visibility of 7.6\%, again consistent with simulations.

Similar to \S\ref{sec: thd}, we were not able to subtract the recorded SCC image due to limited dynamic range in the coronagraphic image. In this case, a bright non-coronagraphic PSF is present in the coronagraphic image due to back-reflections inside the SLM structure, notably at the glass/liquid crystal interface which cannot be anti-reflection coated. To prevent saturation from this ``leakage term,'' the modulation transfer function (MTF) sidelobes in Fig. \ref{fig: eth} (c) are again detected close to the background noise limit. Interestingly, this leakage term is not coherent with the wavefront for which the TGV phase shift is applied, suggesting that the optical path difference from the aforementioned back-reflections is not negligible vs. the coherence length. This is illustrated by (1) the absence of fringes on the leakage term in the coronagraphic image and (2) the absence of the low order features in the MTF sidelobe compared to the central lobe. However, saturating the leakage term in the coronagraphic image would create systematic ringing and aliasing effects in the MTF at the higher spatial frequencies of the fringes (some of which are already seen in Fig. \ref{fig: eth} c), and so algorithmic filtering of this incoherent light is still not an optimal solution to reaching deeper contrasts. Future demonstrations of FAST using liquid crystal technology will implement methods to increase the attenuation of this leakage term, such as solutions proposed by Doelman et al. (2017)\cite{leiden} or Janin-Potiron et al. (2019)\cite{lam_slm}.

\section{Summary}
\label{sec: conclusion}
In G1 and G2 we presented FAST to enable reaching deeper contrasts relative to current exoplanet imaging instruments. In the paper, we further expand on FAST by (1) illustrating the potential scientific gain with both current and future facilities, and (2) presenting the first preliminary laboratory tests of our FAST coronagraphic mask design from G1, both of which are summarized below:
\begin{enumerate}
\item On current 10 m-class telescopes, fast focal plane wavefront sensing, such as FAST, could enable direct imaging and characterization of mature gas giants in reflected light and young exo-Jupiters in thermal emission. On future ELTs, methods like FAST could enable direct imaging of habitable exoplanets. These science cases are enabled by deep contrast gains of more than two orders of magnitude over existing instruments. 
\item In the LESIA laboratories we tested the first FAST focal plane mask designed to boost the SCC fringe visibility to be used in millisecond exposures. Along with additional tests at the ETH laboratories, the measured fringe visibility boost is consistent with simulations. 
\end{enumerate}
Now that we have validated the basic concept of the new coronagraphic mask, we have designed subsequent laboratory demonstrations to mitigate the hardware limitations identified in these first tests in order to demonstrate a deeper contrast gain, which will be discussed and presented in a future paper using our recently assembled high contrast imaging laboratory facilities in Victoria: NRC's Extreme Wavefront lab for the Exoplanet Advanced Research Theme at Herzberg (NEW EARTH). 
\acknowledgments 
 
We gratefully acknowledge research support of the Natural Sciences and Engineering Council of Canada through the Postgraduate Scholarships-Doctoral program, Technologies for Exo-Planetary Science Collaborative Research and Training Experience program, and Discovery Grants Program. The first author thanks LESIA for the accommodation received at Paris Observatory to complete the laboratory tests presented in this paper.

\bibliography{ref} 
\bibliographystyle{spiebib} 

\end{document}